\documentclass[10pt,twocolumn,letterpaper]{article}

\usepackage{iccv}              
\usepackage{colortbl}
\usepackage{graphicx}     
\usepackage{subcaption}  
\usepackage{multirow}

\usepackage{hyperref}

\usepackage{xspace}
\usepackage{newtxtext}

\newtoggle{release}
\togglefalse{release}

\iftoggle{release}{
\newcommand{\roi}[1]{}
\newcommand{\adios}[1]{}
\newcommand{\gleb}[1]{}
\newcommand{\tavi}[1]{}
}
{
\newcommand{\roi}[1]{\textbf{\color{red}[Roi: #1]}}
\newcommand{\adios}[1]{\textbf{\color{magenta}[Yossi: #1]}}
\newcommand{\gleb}[1]{\textbf{\color{orange}[Gleb: #1]}}
\newcommand{\tavi}[1]{\textbf{\color{purple}[Tavi: #1]}}
}

\newcommand{\method}{\textsc{CAFA}\xspace}
\newcommand{\newpara}[1]{\vspace{0.07cm}\noindent \textbf{#1}}

\definecolor{iccvblue}{rgb}{0.21,0.49,0.74}

\def\paperID{5746} 
\def\confName{ICCV}
\def\confYear{2025}

\title{CAFA: a Controllable Automatic Foley Artist}

\author{Roi Benita$^{*+}$\\
Technion\\
{\tt\small roibenita@campus.technion.ac.il}
\and
Michael Finkelson$^{*+}$\\
Hebrew University of Jerusalem\\
{\tt\small michael.finkelson@mail.huji.ac.il}
\and
Tavi Halperin\\
Lightricks\\
{\tt\small tavi@lightricks.com}
\and
Gleb Sterkin\\
Lightricks\\
{\tt\small gsterkin@lightricks.com}
\and
Yossi Adi\\
Hebrew University of Jerusalem\\
{\tt\small yossi.adi@mail.huji.ac.il}
}

\newcommand{\bx}{\mathbf{x}}
\newcommand{\bepsilon}{\mathbf{\epsilon}}
\newcommand{\bz}{\mathbf{z}}
\newcommand{\bI}{\mathbf{I}}
\newcommand{\0}{\mathbf{0}}
\begin{document}

\maketitle
\def\thefootnote{*}\footnotetext{Equal Contribution}
\def\thefootnote{+}\footnotetext{Work done as part of an internship at Lightricks.}
\def\thefootnote{\arabic{footnote}}
\begin{abstract}
Foley is a key element in video production, refers to the process of adding an audio signal to a silent video while ensuring semantic and temporal alignment. In recent years, the rise of personalized content creation and advancements in automatic video-to-audio models have increased the demand for greater user control in the process. One possible approach is to incorporate text to guide audio generation. While supported by existing methods, challenges remain in ensuring compatibility between modalities, particularly when the text introduces additional information or contradicts the sounds naturally inferred from the visuals. In this work, we introduce \method (Controllable Automatic Foley Artist) a video-and-text-to-audio model that generates semantically and temporally aligned audio for a given video, guided by text input. \method is built upon a text-to-audio model and integrates video information through a modality adapter mechanism. By incorporating text, users can refine semantic details and introduce creative variations, guiding the audio synthesis beyond the expected video contextual cues. Experiments show that besides its superior quality in terms of semantic alignment and audio-visual synchronization the proposed method enable high textual controllability as demonstrated in subjective and objective evaluations.\footnote{Samples and code can be found in our \href{https://cafa-vt2a.github.io/CAFA/}{demo page}.}
\end{abstract}
    
\section{Introduction}
\label{sec:intro}
\begin{figure}[t!]
    \centering
    \includegraphics[width=\linewidth]{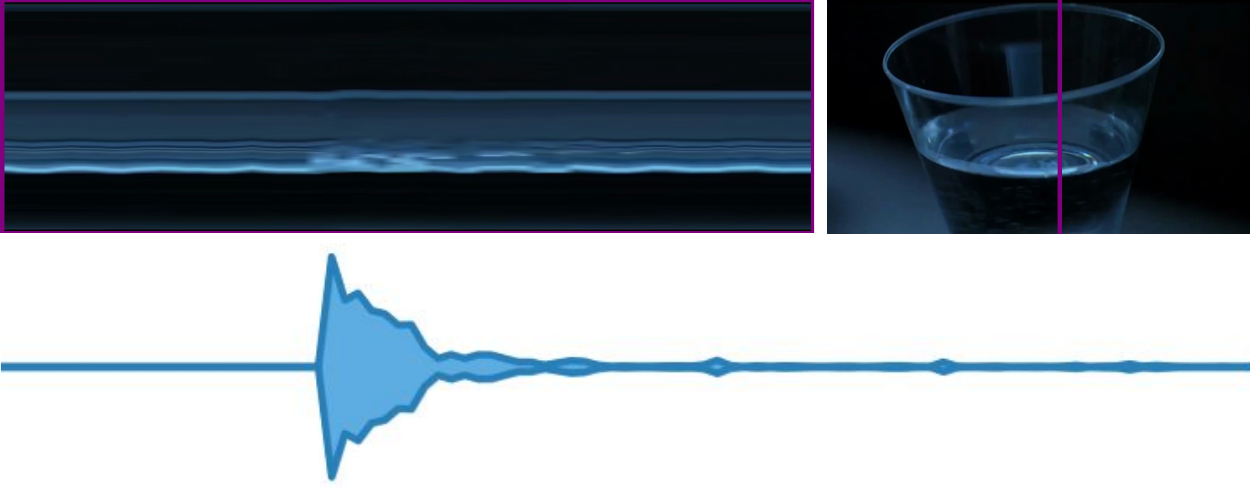}
    \caption{\textbf{Motivation.}
    An iconic scene from \textit{Jurassic Park}, where water in a glass shakes due to the approaching footsteps of a T-Rex. Inferring the generated sound from the video alone is insufficient, as the task is inherently ambiguous. \textbf{Top:} a representative frame and a Y-T slice (from the purple column), where the temporal cue of the shake is faintly visible. \textbf{Bottom:} Our method leverages the prompt "T-Rex Stomping" to generate a synchronized audio track that aligns with both the visual timing and artistic intent.\label{fig:teaser}}
\end{figure}
In recent years, personal content creation has become a major part of everyday life, shaping how we work, entertain, and communicate. One example is \emph{Foley}, the art of adding sound effects to silent videos while ensuring precise semantic and temporal alignment \cite{ament2014foley}. Traditionally, this process was done manually by professional sound designers. However, with the growing demand for fast and immediate personal digital content, the need for automation and accessibility of this process has increased. An effective Foley generation approach should produce high-quality, synchronized audio while also allowing users to creatively shape the sound, balancing precision with creative flexibility.

Building on this need, recent advancements in generative models have led to the development of Video-to-Audio (V2A) models, which aim to automate Foley synthesis and explore cross-modal correspondences \cite{im2wav, iashin2021taming, viertola2024temporally, luo2023diff, wang2025frieren, xing2024seeing}. While these models effectively capture semantic information through global visual representations such as CLIP \cite{radford2021learning}, they often rely on motion-sound relationships for temporal alignment \cite{viertola2024temporally, luo2023diff, wang2025frieren}. Some approaches, including \cite{iashin2021taming}, model motion explicitly using optical flow \cite{horn1981determining}, while others \cite{viertola2024temporally, luo2023diff, wang2025frieren} leverage contrastive learning-based encoders such as CAVP \cite{luo2023diff} and AV-CLIP \cite{iashin2024synchformer} to learn temporally and semantically aligned audio-visual features. Despite these advancements, existing models remain limited to extracting information from the video itself and struggle to incorporate user-provided cues, restricting flexibility and creative control over sound design.

To bridge this gap, Text-and-Video-To-Audio (TV2A) models have been introduced, integrating textual information to enhance control over audio generation \cite{jeong2024read, xing2024seeing, chen2024video, cheng2024taming, liu2024tell, zhang2024foleycrafter}.  By incorporating text, these models allow users to modify audio semantics, add details, and generate diverse sound variations. For instance, text can specify how a sound should be perceived, such as describing a door as creaking or coffee being sipped loudly. Another possibility is introducing creativity through text; a barking dog in a video could instead sound like a meowing cat or a crowing rooster, depending on the accompanying description. In the context of soundtrack design, one would often like to add sounds which do not appear in the video, such as in the iconic scene from \emph{Jurassic Park}, where water in a glass shakes due to the approaching footsteps of a T-Rex; see Figure~\ref{fig:teaser} for a visual example. However, textual conditioning is often not sufficiently strong or may come at the expense of temporal alignment between video and audio. Additionally, when the text describes semantics that differ from the video, existing models frequently struggle to generate a natural and coherent audio signal (See Section~\ref{sec:results}).

Various methods have been explored for integrating text into multimodal systems. A common strategy involves jointly training video, text, and audio representations to capture shared semantics \cite{cheng2024taming, chen2024video}. However, this requires retraining the entire network whenever modifications are made, leading to high computational costs. Alternatively, a training-free method \cite{xing2024seeing} leverages a shared latent space to link the modalities, eliminating the need for retraining. Yet, this introduces test-time optimization, increasing inference time and potentially degrading output quality and alignment. A middle-ground solution employs a modality adapter (e.g., ControlNet mechanism \cite{zhang2023adding}), which uses video inputs to condition a pretrained Text-to-Audio (T2A) model \cite{jeong2024read, zhang2024foleycrafter}, providing an effective way to incorporate video information into text-driven audio synthesis.

In this work, we introduce \method, which stands for Controllable Automatic Foley Artist, a novel text-and-video-to-audio model that extends beyond temporal and semantic synchronization, allowing users to shape and control sound through textual cues. \method leverages a ControlNet like modality adapter to flexibly integrate pretrained T2A models with video-based features while maintaining a relatively low training cost ($48$ A$100$ GPU hours for \method vs. $304$ H$100$ GPU hours for the baseline method). Specifically, we explore Stable-Audio-Open \cite{evans2024stable} and TangoFlux \cite{hung2024tangoflux} as T2A models. To extract temporal and semantic features, we experiment with AVCLIP \cite{iashin2024synchformer} and CLIP \cite{radford2021learning} as the video representations. \method achieves high-quality audio synthesis, temporal synchronization, and contextual alignment performance comparable to state-of-the-art V2A and TV2A models. Additionally, it significantly surpasses existing TV2A approaches when the text and visual conditioning cues are semantically different, demonstrating greater control over generated sound.

Our main contributions are:
(i) We introduce \method, a novel TV2A model that allows the generation of temporally and semantically aligned audio while providing extensive textual control over the generated audio; (ii) We evaluate \method against existing V2A and TV2A models, demonstrating comparable performance in audio quality and video-audio compatibility, while achieving superior performance for textual control, as validated through disentanglement experiments, objective evaluations, and human studies; (iii) \method is built on the modality adaptation (via a ControlNet mechanisem), enabling precise temporal control while offering a versatile framework that supports modular integration, accommodating different T2A models (Stable Audio Open and TangoFlux). Additionally, it facilitates the efficient incorporation of video representations, leading to more effective training compared to alternative methods.

\begin{figure*}[t!]
    \centering
    \begin{subfigure}{0.73\textwidth}
        \includegraphics[width=\textwidth]{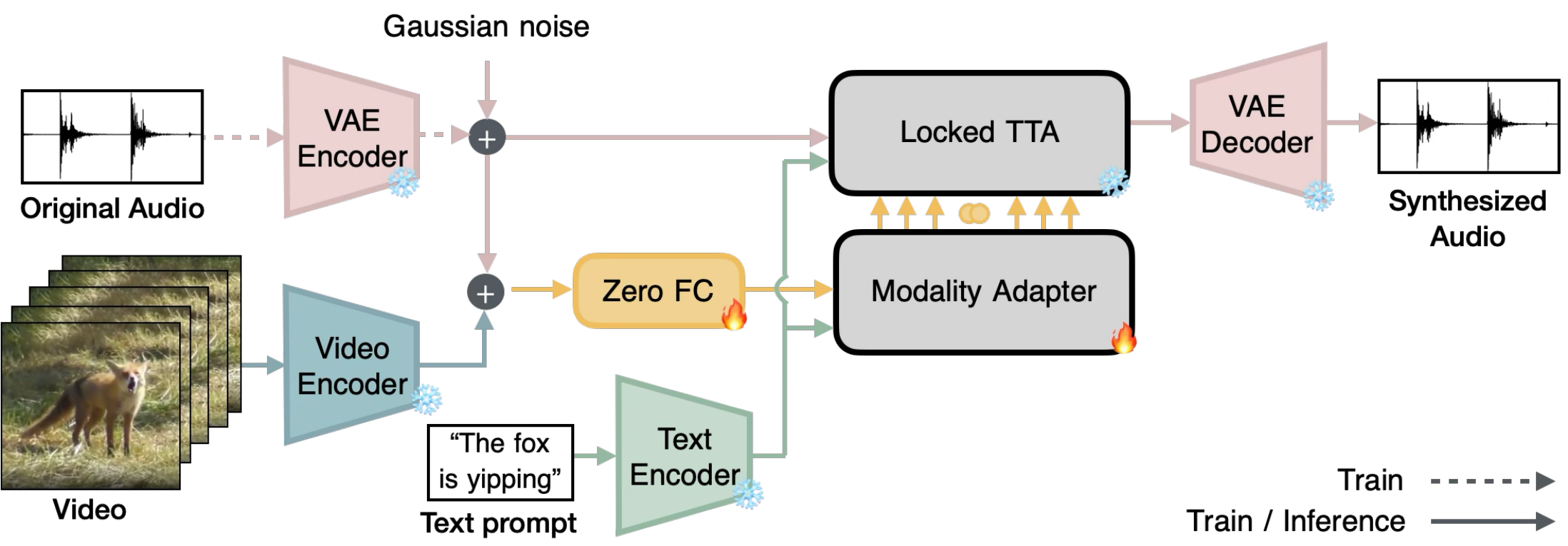}
        \caption{}
        \label{fig:ZoomOut_Arch}
   \end{subfigure}
   \hfill
    \begin{subfigure}{0.23\textwidth}
        \includegraphics[width=\textwidth]{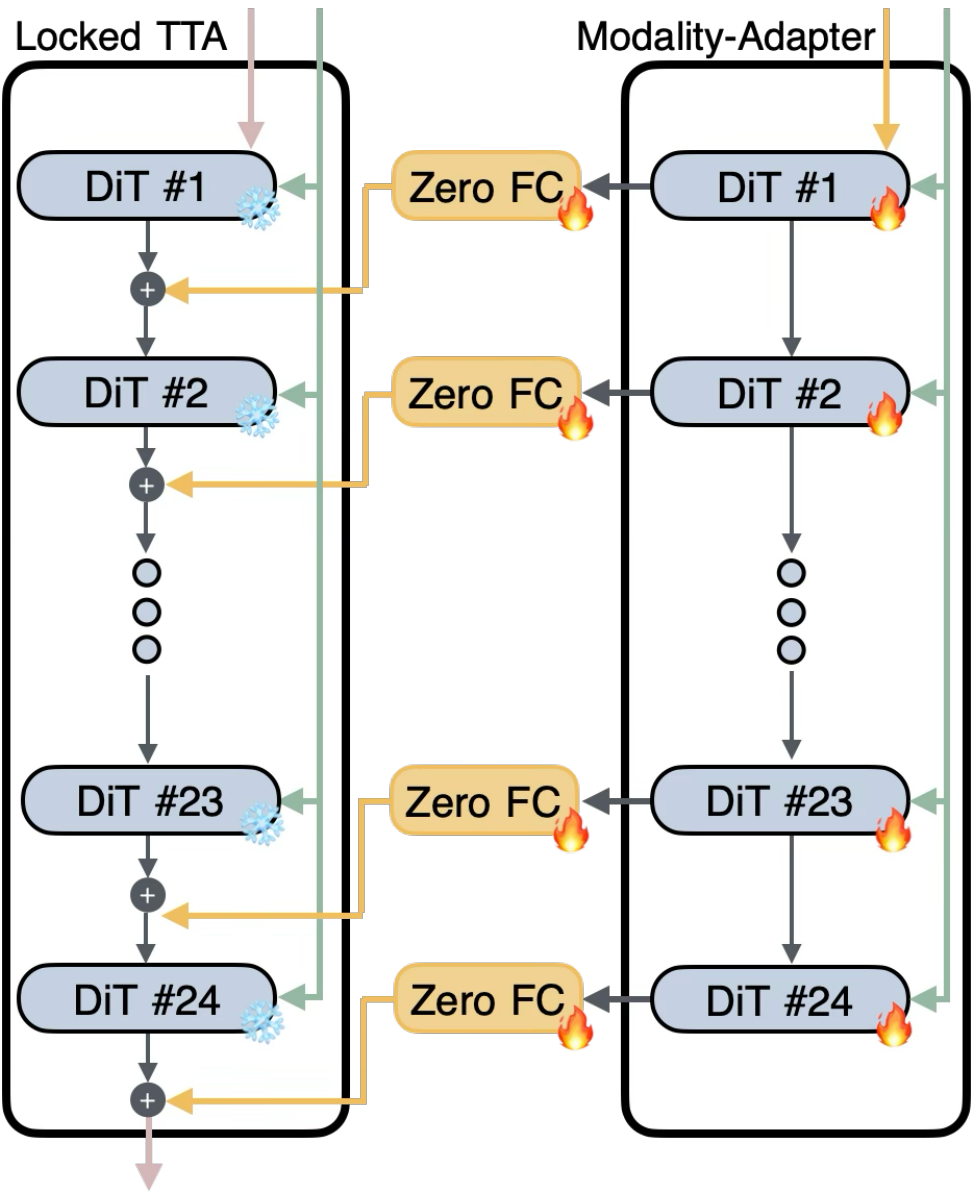}
        \caption{}
        \label{fig:ZoomIn_Arch}
    \end{subfigure}
     
    \caption{(a) \textbf{Method overview:} our model is text-and-video-to-audio, leverages pretrained models for audio generation, and video encoding. The original audio and VAE audio encoder are only used during training. (b) \textbf{Adaptor:} Illustration of the internal connectivity between the backbone T2A model and our video conditioning adaptor,  with fully connected (FC) layers explicitly shown.}
    \label{fig:CAFA_Arch}
\end{figure*}

\section{Background}
\label{sec:Background}

\subsection{Latent Diffusion Models}

Latent Diffusion Models (LDMs)~\cite{rombach2022high} are a class of generative models that perform a diffusion process within a learned latent space $ \bz $. A Variational Autoencoder (VAE) encodes a data sample $ \bx \sim p(\bx) $ into a lower-dimensional latent space $ \bz \in \mathbb{R}^d $ using an encoder $ \mathcal{E} $, while a decoder $ \mathcal{D} $ reconstructs $ \bx $. Performing diffusion in this reduced space significantly reduces computational cost while maintaining high-quality generation.

The diffusion process follows two Markovian paths: the forward and reverse processes. In the forward process, a clean latent representation \( \bz_0 \) is gradually corrupted with additive Gaussian noise:  
\begin{equation}
\bz_t = \sqrt{\alpha_t} \bz_{t-1} + \sqrt{1 - \alpha_t} \bepsilon_t,
\end{equation}
where $ \left\{\alpha_t\right\}_{t=1}^T $ defines the noise schedule, $ \bepsilon_t \sim \mathcal{N}(\mathbf{0}, \mathbf{I}) $ and $ \bz_T \sim \mathcal{N}(\mathbf{0}, \mathbf{I}) $.
A key consequence of the forward process is its marginal distribution:
\begin{equation}
\bz_t = \sqrt{\bar{\alpha}_t} \bz_0 + \sqrt{1 - \bar{\alpha}_t} \bepsilon, \quad \bepsilon \sim \mathcal{N}(\mathbf{0}, \mathbf{I}),
\end{equation}
where $\bar{\alpha}_t=\prod_{t=1}^{T}\alpha_t$. A neural network, trained as a denoiser, learns to estimate $\bepsilon$ given the noisy input $ \bz_t $, the timestep $t$ , and conditioning information $ c $, such as encoded text. The training objective minimizes the difference between the true noise and the predicted noise:
\begin{equation}
\mathcal{L} = \mathbb{E}_{\bz, \bepsilon \sim \mathcal{N}(\0,\bI), t, c} \left[ \| \bepsilon - \bepsilon_{\theta} (\bz_t, t, c) \| \right].
\end{equation}

By using the network output, the reverse process aims to reconstructs \( \bz_0 \) from \( \bz_T \) by iteratively denoising it. While initial works \cite{ho2020denoising, song2020denoising} formulated this process discretely, \citet{song2020score} showed that it can be equivalently expressed as an Ordinary Differential Equation (ODE) which can be solved numerically using dedicated solvers. Specifically, Stable Audio Open \cite{evans2024stable} employs DPM-Solver++ \cite{lu2022dpm} and follows the the v-objective approach \cite{salimans2022progressive}.

\newpara{Classifier-Free-Guidance (CFG)} is a widely used method to improve performance in conditional generative models, originally demonstrated in diffusion-based image generation approaches~\cite{ho2022classifier}. CFG is an effective control mechanism for steering the inference process to better align with provided conditioning signals. Specifically, it modifies the predicted noise by linearly combining the estimates from a conditional diffusion model and a jointly trained unconditional model, resulting in the following formulation:
\begin{equation}
\tilde{\bepsilon}_{\theta}(\bz_t, c, t) = \bepsilon_{\theta}(\bz_t, t) + \gamma \left(\bepsilon_{\theta}(\bz_t, t, c) - \bepsilon_{\theta}(\bz_t, t) \right)
\end{equation}

where $\gamma$ determines the strength of the guidance, with higher values enforcing stronger adherence to the conditioning signal.

\subsection{ControlNet Mechanism}
The ControlNet mechanism was initially introduced as a neural network architecture for controlling text-to-image models through spatially localized conditioning (e.g., canny edge and depth maps)\cite{zhang2023adding}. It preserves the quality and stability of a large pretrained model by locking its weights, while enabling the incorporation of control signals through a replicated copy of that backbone model. These components are connected via zero-initialized convolutional layers, allowing a gradual integration which minimizes noise interference during training.

\section{Method}
\label{sec:Method}

\method consists of two main components: a pretrained backbone Text-To-Audio (T2A) model, with frozen weights to maintain audio quality, and a trainable modality adapter that integrates temporal and semantic video information. The components are linked through Zero Fully-Connected (FC) Layers, which prevent noise from disrupting the backbone model during early training. This structure allows us to benefit from the long pre-training of the foundational T2A model instead of training all three modalities from scratch. Figure \ref{fig:ZoomOut_Arch} provides a high-level overview of the proposed method.

\newpara{Text-to-Audio Backbone.} A variational autoencoder~\cite{evans2024stable} encodes the input signal $\bx \in \mathbf{R}^{2 \times L}$ (2 channels for stereo) into a latent representation $\bz \in \mathbf{R}^{T \times C}$, with $L$ denoting the temporal length of the audio, while $T$ and the $C = 64$ correspond to the temporal dimension and feature size, respectively. Next, noise is added to the latent representation, producing $\bz_t$, which is then processed by a core architecture built from a stack of Diffusion Transformer (DiT) blocks~\cite{peebles2023scalable}, with model-dependent variations. Each DiT block is controlled by a text input, encoded by a pretrained text encoder, guiding the generation process. Furthermore, a timing mechanism sets the signal length and fills the rest with silence.

We experiment with two T2A models: Stable Audio Open \cite{evans2024stable} and TangoFlux \cite{hung2024tangoflux}. These models provide the underlying architecture for creating high-quality stereo audio content at a sampling rate of $44.1$ kHz based on textual descriptions. Although the architectural designs and sampling methods differ, both models follow a comparable core structure. This allows us to evaluate the flexibility of our approach, demonstrating that it is not tied to a single family of models and can be applied across various T2A models.

\newpara{Modality Adapter.} The adapter is tasked with incorporating the video to guide the T2A model generation. Our design is inspired by 
ControlNet~\cite{zhang2023adding}, with some notable differences. First, it operates on the temporal domain, requires synchronization of features from different modalities. Second, our T2A model is a DiT, rather than a U-Net \cite{ronneberger2015u}, changing the connectivity between the adapter and the base model. Specifically, after the preprocessing stage, informative features from the video, $ E_v \in \mathcal{R}^{T \times C} $, are passed through a Zero FC layer and added to $ \bz_t $. Additionally, the hidden states, extracted from each DiT block, are processed through Zero FC layers and added to the backbone model, as depicted in Figure \ref{fig:ZoomIn_Arch}. Only the adapter is being trained, while the T2A, Text Encoder, and Video Encoder, all kept frozen.

\newpara{Video Representation.}  We experiment with AVCLIP \cite{iashin2024synchformer} and CLIP \cite{radford2021learning}, both of which are trained with contrastive learning. AVCLIP processes $0.64$-second video-audio segments and applies InfoNCE loss \cite{oord2018representation} to differentiate between positive and negative samples. Its video encoder, built on MotionFormer \cite{patrick2021keeping}, enhances the modeling of dynamic scenes by capturing implicit motion paths. To ensure better alignment with $\bz_t$, we increased the segment overlap compared to the original work, where $216$ samples roughly correspond to $10$ seconds.
We further examine the contribution of CLIP to the semantic scene understanding, similarly to MMAudio~\cite{cheng2024taming}. Since CLIP is less sensitive to object locations \cite{chen2024contrastive}, we introduce a temporal dimension to capture significant visual changes. We extract CLIP representations at 5 FPS and interpolate to match the temporal dimension of $\bz_t$. Finally, both representations are preprocessed before being integrated with the modality adapter, as detailed in the supplementary~\cite{Authors}.

\newpara{Asymmetric Classifier-Free Guidance.} Incorporating CFG into our model with text as the conditioning signal $c$ presents challenges in effectively utilizing video conditioning at higher guidance scales. To address this limitation, we propose \emph{Asymmetric Classifier-Free Guidance}, where the modality adapter's output is selectively modulated in the conditional and unconditional pathways during synthesis. Unlike the standard approach, which equally integrates 
the modality adapter into the backbone model in both pathways, our method introduces an asymmetric scaling factor, $ 0 \leq \alpha \leq 1 $, reducing the influence of the adapter in the unconditional path,
\begin{equation}
    h^{'}_{i,c} = h_{i,c} \quad , \quad h^{'}_{i,uc} = \alpha \cdot h_{i,uc},
\end{equation}
where 
$ h_i = [ h_{i,c}, h_{i,uc}] $ are the hidden states, with $ h_{i,c} $ representing the conditional path and $ h_{i,uc} $ the unconditional path. 
Consequently, under $ \alpha < 1 $, this adjustment induced controlled disparity between $ \bepsilon_{\theta}(\bz_t, t, c) $ and $ \bepsilon_{\theta}(\bz_t, t) $ effectively amplifies the video conditioning signal. Standard CFG is a special case, corresponding to $ \alpha = 1 $. Our experiments demonstrate that this simple yet effective modification significantly enhances adherence to video conditioning while maintaining high generation quality and text controllability.

\section{Experimental Setup}
\label{sec:Experiments}

\subsection{Datasets}

The proposed model was trained using two datasets: VGGSound~\cite{chen2020vggsoundlargescaleaudiovisualdataset} and VisualSound \cite{viertola2024temporally}. VisualSound is a subset of VGGSound filtered to include samples with high ImageBind \cite{Girdhar2023ImageBindOE} scores. Both datasets contain $10$-second video clips across diverse acoustic categories accompanied by video captions. For TV2A and V2A evaluation, we use the VGGSound dataset and  VGGSound-Sparse~\cite{sparse2022iashin}, which is a subset of VGGSound containing $12$ categories of naturally sparse audio events such as ``dog barking'' or ``playing tennis''. 

\subsection{Baseline Methods}
We compare \method against several state-of-the-art models, namely MMAudio~\cite{cheng2024taming} (\textit{large\_44k\_v2} version), FoleyCrafter~\cite{zhang2024foleycrafter}, VATT~\cite{liu2024tell}, ReWaS~\cite{jeong2024read}, Frieren~\cite{wang2025frieren}, and MultiFoley~\cite{chen2024multifoley}. For FolyCrafter, we follow original configurations by formatting text prompts as ``The sound of \texttt{<label>}''. For VATT, Frieren, and MultiFoley, we consider the samples provided by the respective authors. MultiFoley samples were only available for a subset of the test set with high ImageBind scores. ReWaS is evaluated using its default configuration. Notice, with $5$-second samples, unlike other models that produce $8$-second samples. Hence, for a fair comparison against this model we truncate the videos to $5$-second. 

\subsection{Implementation Details}
\method models are initially trained for $48$k steps on VGGSound, followed by fine-tuning for $33$k more steps on VisualSound. Training was performed with a batch size of $16$, using the AdamW \cite{loshchilov2017decoupled} optimizer, on a single A$100$ GPU. We generate samples using CFG = $7$, Asymmetric CFG scale of $\alpha=0.5$, and $50$ inference steps, while keeping the rest of the TTA model configuration unchanged. Our model is trained on $10$-second samples, and the output truncated to $8$ seconds for fair comparison with the baseline methods. 

\subsection{Evaluation Metrics}
We evaluate model performance across four complementary dimensions that capture different aspects of audio-visual generation: Audio Quality, Audio-Visual Semantic Alignment, Audio-Visual Temporal Alignment, and Audio-Text Semantic Alignment.

\newpara{Audio Quality.} We employ three established metrics to assess the fidelity and naturalness of generated audio: (i) Fréchet Audio Distance (FAD)~\cite{kilgour2019frechetaudiodistancemetric} which measures distributional similarity between features extracted from ground truth and generated audio; (ii) Kullback-Leibler Distance (KL)~\cite{Kullback1951OnIA}, which quantifies the difference between probability distributions of per-sample ground truth and generated audio features; and (iii) Inception Score (IS)~\cite{Salimans2016ImprovedTF} that evaluates the generated audio quality independently of ground truth references.
We utilize PANNS~\cite{kong2020pannslargescalepretrainedaudio} as the features extractor for all three audio quality metrics.

\newpara{Audio-Visual Semantic Alignment.} We leverage ImageBind (IB)~\cite{Girdhar2023ImageBindOE} to quantify semantic similarity between the ground truth video and generated audio. This cross-modal embedding model measures whether the generated audio contains appropriate sounds for the visual content.

\newpara{Audio-Visual Temporal Alignment.} We utilize DeSync~\cite{cheng2024taming} (also known as Sync~\cite{viertola2024temporally} or AV-Sync~\cite{chen2024multifoley}) to measure temporal synchronization between audio and video. DeSync calculates the average absolute offset (in seconds) between ground truth video and generated audio using Synchformer~\cite{iashin2024synchformer} predictions. Following prior work\cite{cheng2024taming}, we average DeSync scores from the first and last $4.8$ seconds of audio to accommodate Synchformer's limited context window.

\newpara{Audio-Text Semantic Alignment.} We employ CLAP~\cite{laionclap2023} to evaluate similarity between generated audio and textual descriptions of the video by calculating the cosine similarity between the text and audio embeddings.

\begin{table}
\centering
\resizebox{0.95\linewidth}{!}{ 
\begin{tabular}{l ccccc}
\toprule
\bf Model & \bf FAD↓ & \bf IS↑ & \bf CLAP↑ & \bf Acc↑ & \bf DeSync↓ \\
\midrule
FC & 57.00 & \underline{6.10} & 0.10 & 0.69 & 1.30 \\
MMA & \textbf{16.43} & 6.77 & 0.10 & 0.36 & \textbf{0.57} \\
ReWaS & 38.94 & 5.13 & 0.09 & 0.74 & 1.19 \\
\method (Ours) & \underline{27.33} & 5.63 & \textbf{0.21} & \textbf{0.87} & \underline{0.81} \\
\bottomrule
\end{tabular}}
\caption{\textbf{Semantically different text and video conditioning.} Our method surpasses strong concurrent SOTA in terms of prompt adherence by a large margin. Arrows indicate whether higher (↑) or lower (↓) values are better. FC: FoleyCrafter, MMA: MMAudio. 
\label{tab:disent}}
\end{table}

\section{Results}
\label{sec:results}
We start by evaluating model performance when considering semantically different text and video conditioning. Ideally, we expect the model to generate textually described audio aligned with the visual cues. For that we generate audio for each video in VGGSound-Sparse, using the ground truth video paired with captions from each of the $11$ other categories. This cross-category approach creates a challenging scenario where models must follow textual instructions that deliberately conflict with visual content.

Notice, under this setup we compute the CLAP similarity score between the new caption and generated audio. We also use CLAP as a classifier between the new caption and the GT caption for the generated audio, reporting binary classification accuracy (Acc). 

For FoleyCrafter, we follow \cite{chen2024multifoley} and disable the semantic adapter to allow the model to generate the requested caption, using only the temporal adapter to retrieve information from the video.

\begin{figure*}[t!]
    \centering    
    \includegraphics[width=\linewidth]{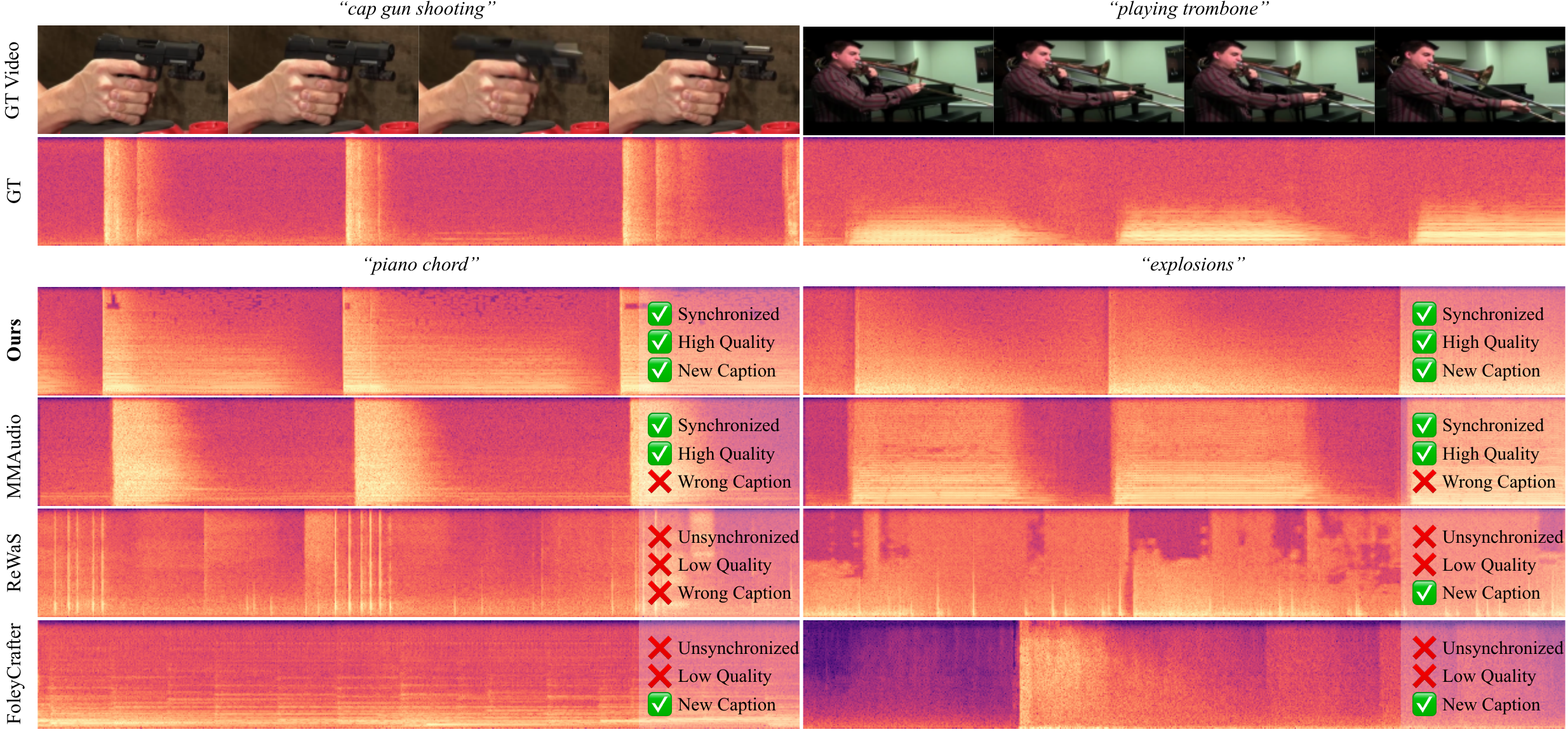}
    \caption{\textbf{Qualitative Comparison of Text-Video Disentanglement.} A comparative analysis of various TV2A models: Ground Truth (GT), \method (ours), MMAudio, ReWaS, and FoleyCrafter, using the same configurations as in Table~\ref{tab:disent}. Our model consistently delivers synchronized, high-quality generations that accurately adhere to the requested target captions, outperforming other approaches. Full videos presented at \href{https://anonymous.4open.science/w/CAFA} {\textit{our demo page}}.\label{fig:new captions comparison}}    
\end{figure*}

Results are summarized in Table~\ref{tab:disent}, with visual examples depicted in Figure~\ref{fig:new captions comparison}. MMAudio achieves the highest audio quality (best FAD and IS scores) and temporal alignment (DeSync), but critically fails at semantic control. Its Acc scores of $0.58$ (w\textbackslash negp) and $0.36$ (w\textbackslash o negp) demonstrate that it systematically generates audio corresponding to the visual content rather than adhering to the requested text prompt, essentially negating the purpose of text-guided generation. Despite disabling FoleyCrafter's semantic adapter specifically for this experiment, it exhibits severe temporal misalignment (worst DeSync score of $1.30$) producing outputs that fail to synchronize with visual events. This renders its generation ineffective even for basic audio-visual correspondence. ReWaS performs sub-optimally across all evaluation dimensions—producing lower audio quality, weaker temporal alignment, and poorer semantic control compared to our approach—without exhibiting a particular strength in any area to compensate for these deficiencies. In contrast, \method successfully balances all critical requirements: achieving strong audio quality (second-best FAD and comparable IS), effective temporal alignment (second-best DeSync), while significantly outperforming all competitors in semantic controllability with a CLAP score of $0.21$ (compared to $<0.1$ for all others) and Acc of $0.87$. This comprehensive performance profile makes our approach uniquely capable of producing high-fidelity, temporally-aligned audio that accurately follows semantic text instructions.

\begin{table}
    \centering
    \resizebox{\linewidth}{!}{    
    \begin{tabular}{l ccc c c c}
        \toprule
        \bf Model &\bf  FAD↓ &\bf  KL↓ &\bf  IS↑ &\bf  IB↑ &\bf  DeSync↓ &\bf  CLAP↑ \\
        \midrule
        FC$\dagger$  & 13.68 & 2.56 & 10.68 & \underline{0.27} & 1.30 & \underline{0.12} \\
        MMA$\dagger$ & \phantom{0} \textbf{5.32} & \textbf{1.64} & \textbf{17.18} & \textbf{0.33} & \textbf{0.77} & \textbf{0.23} \\
        Frieren$\dagger$  & \underline{11.76} & 2.70 & 12.33 & 0.23 & 1.04 & 0.11 \\
        \midrule
        FC & 22.17 & 2.87 & 13.30 & 0.16 & 1.31 & 0.18 \\
        MMA & \phantom{0}\textbf{6.89} & \underline{1.65} & \textbf{20.44} & \textbf{0.34} & \textbf{0.76} & \textbf{0.25} \\
        VATT & \underline{11.13} & \textbf{1.48} & 11.85 & 0.25 & 1.28 & 0.15 \\
        ReWaS* & 14.71 & 2.69 & \phantom{0}8.45 & 0.15 & 1.18 & 0.18 \\
        MF* & 13.51 & \underline{1.65} & \underline{15.89} & \underline{0.27} & 1.04 & \underline{0.23} \\                
         \method (Ours) & 12.60 & 2.02 & 13.45 & 0.21 & \underline{0.96} & \underline{0.23} \\               
        \bottomrule
    \end{tabular}}
    \caption{\textbf{Quantitive comparison.} We report results comparing standard V2A models, V2A variants of TV2A models (indicated by $\dagger$), TV2A models, and our method. FC:FoleyCrafter, MMA:MMAudio, MF:MultiFoley. $*$ indicates variations - we compare with ReWaS on samples trimmed to $5$ seconds, and compare with MultiFoley on their selected subset of the test set. \label{tab:results}}
\end{table}

\newpara{Semantically Aligned Visual and Textual Conditions.} Next, we compare \method where the visual and textual conditions are semantically aligned. We compare the proposed method against V2A and VT2A methods, considering the VGGSound test set using the standard configurations. For V2A models, we do not use textual descriptions. Results are presented in Table~\ref{tab:results}. 

While MMAudio emerges as the strongest performer, comparison of MMAudio with and without text conditions reveals minimal benefits from textual inputs. Specifically, FAD got slightly worse when text is added, while other metrics remain largely unchanged. Interestingly, a similar pattern emerges in FoleyCrafter, which shows mixed performance with text conditioning - its FAD, KL, IB got worsen, while IS and CLAP improved, and DeSync stays unchanged. These results strengthens our findings in Table~\ref{tab:disent}, demonstrating the improved flexibility and controllability of the proposed method compared to the baseline methods.

\method demonstrates balanced performance across all evaluation metrics, achieving the second-best scores in both DeSync and CLAP while maintaining competitive audio quality metrics. These results indicate effective multi-modal conditioning without compromising performance in any single dimension.

\begin{table}[t!]
    \centering
     \resizebox{0.9\linewidth}{!}{ 
     \begin{tabular}{l l c c}
        \toprule
        \textbf{Comparison} & \textbf{Criterion} & \textbf{\%(OC)} & \textbf{\%(RC)} \\
        \midrule
        \multirow{3}{*}{\method vs MMAudio} 
            & Align.          & 0.24 & 0.24 \\
            & Quali.            & 0.29 & 0.26 \\
            & PA   & 0.39 & 0.87\\
        \midrule
        \multirow{3}{*}{\method vs FoleyCrafter} 
            & Align.          & 0.83 & 0.95 \\
            & Quali.            & 0.81 & 0.88 \\
            & PA   & 0.82 & 0.69 \\
        \midrule
        \multirow{3}{*}{\method vs ReWas} 
            & Align.          & 0.96 & 0.96 \\
            & Quali.            & 0.95 & 0.89 \\
            & PA   & 0.94 & 0.77 \\
        \bottomrule
    \end{tabular}}
    \caption{\textbf{Human study.} Win rate results ($\%$) of \method against three baseline methods. Results are reported for: time alignment (Align.), audio quality (Quali.), and prompt adherence (PA), considering the original caption (OC) and unaligned caption (UC).\label{tab:audio_winrate}}
\end{table}

\newpara{Human Study.} Lastly, we evaluate the subjective quality of audio generated by the proposed method. We conduct a user study using the following evaluation protocol considering both the original textual captions (OC) and visually unaligned text caption (UC). We follow the above mention protocol. 

Each participant was presented with a pair of videos side by side. For every pair, the participant was asked three distinct questions: (i) Prompt Adherence (PA): \emph{Which audio better matches the description ‘[text prompt]’?} This question measured how well the audio corresponds to the prompt; (ii) Alignment: \emph{Which audio better aligns with the timing of visual movements and events in the video?'} This question assessed the synchrony between audio events and the corresponding visual actions; (iii) Quality: \emph{Which audio has higher overall technical quality (considering naturalness, clarity, and lack of artifacts)?} This evaluates the technical fidelity of the audio signal. Participants provided independent responses for each of these questions. In both protocols, the ordering of the video pairs was randomized to mitigate any potential bias. The protocol was implemented using a custom web interface built on Amazon SageMaker Ground Truth. We use $17$ videos from VGGSound test set to generate two audio samples: with an original caption and a new one by every method. Then each pair of \method vs competitor (in random order) was shown to $6$ raters and the results were averaged across $6$ annotations per pair.

Win rate results of the proposed method against the evaluated baselines are presented in Table~\ref{tab:audio_winrate}. When comparing to FoleyCrafter and ReWas the proposed method achieves superior performance across all setups. As expected, when comparing to MMAudio, the proposed method reach inferior performance considering both audio quality and time alignment, however, under the unaligned text and visual conditioning, \method reach significantly better performance than MMAudio. 

\begin{table}[t!]    
    \centering
     \resizebox{\linewidth}{!}{
     \begin{tabular}{l ccc c c c}
        \toprule
        & \multicolumn{3}{c}{\bf Audio Quality} & \multicolumn{1}{c}{\bf A-V Sem.} & \multicolumn{1}{c}{\bf A-V Tem.} & \multicolumn{1}{c}{A-T} \\
        \cmidrule(lr){2-4} \cmidrule(lr){5-5} \cmidrule(lr){6-6} \cmidrule(lr){7-7}
        \bf Model & \bf FAD↓ & \bf KL↓ & \bf IS↑ & \bf IB↑ & \bf DeSync↓ & \bf CLAP↑ \\
        \midrule
        \method-B & \textbf{12.57} & 2.04 & 11.84 & \underline{0.21} & \underline{1.00} & \textbf{0.23} \\
        \method & \underline{12.60} & \underline{2.02} & 13.45 & \underline{0.21} & \textbf{0.96} &\textbf{0.23} \\
        \method-C & 14.44 & \textbf{1.98} & \underline{14.18} & \textbf{0.22} & 1.02 &\textbf{0.23} \\
        \method-TF & 19.94 & 2.16 & \textbf{16.94} & 0.20 & 1.12 & \textbf{0.23} \\
        \bottomrule
    \end{tabular}}
    \caption{\textbf{Model ablation.} We compare different variants of our model, and argue that our default model \method strikes an overall good balance between audio quality, and semantic and temporal alignment to video and prompt. \method-B - our model before finetuning on VisualSound. \method-C - additionally leverage CLIP visual features. \method-TF - use TangoFlux\cite{hung2024tangoflux} as base T2A model. \label{tab:model variation ablation}} 
\end{table}

\section{Analysis}
\label{sec:Ablation}

\newpara{Model Variations.} We first evaluate several architectural variants of \method. Specifically, we consider: (i) \method-B, which is identical to \method but does not include additional finetuning phase on VisualSound, allowing us to evaluate training efficiency; (ii) \method-C, which leverages both AV-CLIP and CLIP as visual encoders, combined via MLP.  This model was trained for $84$k steps on VGGSound train split with a batch size of $8$ and the same optimizer as \method; and (iii) \method-TF, that leverages TangoFlux as base T2A model with AV-CLIP visual encode.  \method-TF was trained for $32$k steps on VGGSound train split with batch size $64$ and the same optimizer as \method. For variants using StableAudio-Open, we used a CFG value of $7.0$, asymmetric CFG $\alpha=0.5$, and $50$ inference steps. For \method-TF, we used a CFG value of $4.5$, asymmetric CFG $\alpha=0.8$, and $50$ steps.

Results, presented in Table~\ref{tab:model variation ablation}, shows that all variants achieve similar performance across metrics, with minimal differences in FAD, IS, KL, and DeSync. The \method-C results demonstrate that adding CLIP visual conditioning alongside AV-CLIP provides no meaningful benefits, highlighting that the combination of AV-CLIP with text conditioning is sufficient for effective audio generation. The \method-B variant, trained for only $48$k steps, performs slightly worse than \method. Finally, the performance of \method-TF confirms that our method generalizes effectively across different T2A base models.

\paragraph{Training Efficiency Analysis.} As shown in Table~\ref{tab:training_cost} in the Appendix, \method demonstrates significant training efficiency compared to other state-of-the-art models. While direct comparison is challenging due to differences in reporting methodologies, hardware configurations, and training approaches, several observations can be made. \method requires substantially fewer training steps ($81$k total) compared to models like Frieren ($2.4$M steps) and MultiFoley ($650$k steps). Even our base model (\method-B), trained for only $48$k steps, achieves performance comparable to models trained for much longer periods. The adapter-based approach allows \method to leverage pre-trained text-to-audio models effectively, reducing the need for extensive training from scratch. While MMAudio reports $304$ GPU hours on H$100$ hardware, \method estimated $48$ GPU hours on A$100$ hardware represents a more efficient use of compute resources when considering the relative performance. The modular architecture of \method facilitates efficient training while maintaining high performance across audio quality, temporal alignment, and textual control metrics. This efficiency analysis further highlights the practical advantages of our approach, making \method more accessible for research and potential real-world applications.

\begin{figure}[t!]
    \centering
    \includegraphics[width=1\linewidth]{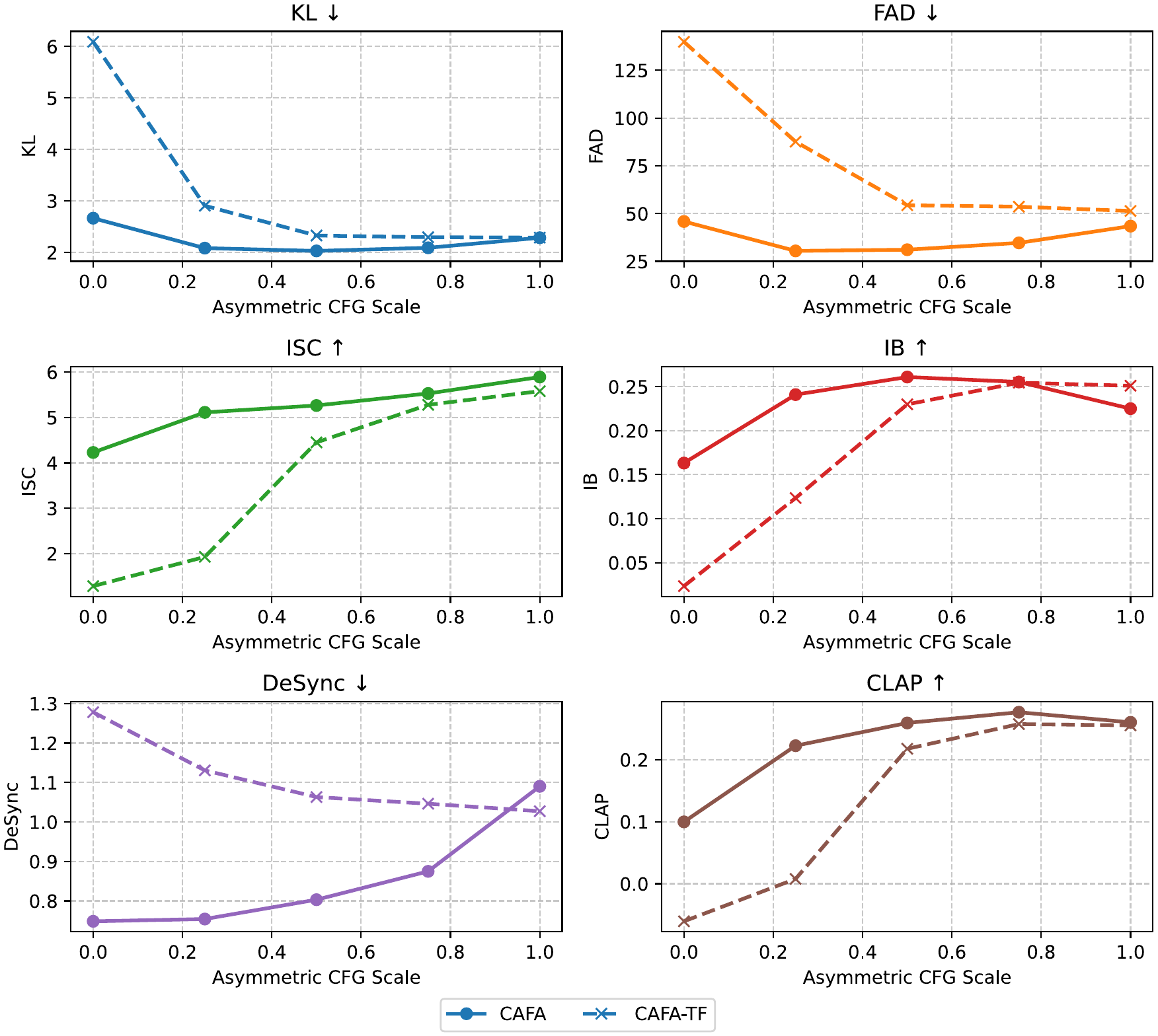}
\caption{Comparison of Asymmetric CFG Scaling Values. CAFA-TF is our adapter applied with TangoFlux\cite{hung2024tangoflux}, while the default implementation uses StableAudio-Open\cite{evans2024stable}.\label{fig:cfg ablation}}
\end{figure}

\newpara{Asymmetric CFG Scaling.} Finally, we investigate different settings for asymmetric CFG scaling. We evaluate both \method and \method-TF over the VGGSound-Sparse test set with the same configurations as described at Section \ref{sec:results}. Figure~\ref{fig:cfg ablation} depicts the results, illustrating the trade-off between different scaling parameters. Our analysis shows that for both models, the trends are similar, with values between $0.5$ and $1$ corresponding to the best results.
\section{Related work}
\label{sec:Related_Work}
\newpara{Text To Audio Models.}
The field of Text-to-Audio (T2A) has advanced significantly, with ongoing improvements in text representations and high-quality audio generation. Early approaches included AudioGen \cite{kreuk2022audiogen}, an autoregressive model employing a discrete waveform representation, while DiffSound \cite{yang2023diffsound} adopted discrete diffusion, removing the need for autoregressive token decoding. As the field evolved, models like AudioLDM \cite{liu2023audioldm}, StableAudio 1\cite{evans2024fast}, and Make-An-Audio \cite{huang2023make} leveraged latent diffusion and incorporated CLAP \cite{laionclap2023} embeddings to improve text decoding. Recognizing the importance of capturing temporal, acoustic, and semantic information, newer models such as Make-An-Audio 2 \cite{huang2023make}, AudioLDM 2\cite{liu2024audioldm}, and Tango \cite{ghosal2023text} integrated large language models (LLMs) \cite{raffel2020exploring} to enhance text-audio alignment. Building on these advancements, Tango 2 \cite{majumder2024tango} further refined temporal alignment by employing Direct Preference Optimization (DPO) \cite{rafailov2023direct}.

Our work is built on StableAudio Open \cite{evans2024stable} and TangoFlux \cite{hung2024tangoflux}, designed for high-quality text-to-audio generation. StableAudio Open is a latent diffusion model that generates stereo audio up to 47 seconds from text input, utilizing a T5 text encoder for text processing and enabling control over output length. In contrast, TangoFlux is based on rectified flow, producing stereo audio up to 30 seconds at 44.1 kHz, while leveraging a pretrained autoencoder from StableAudio Open to enhance efficiency. Additionally, TangoFlux incorporates CLAP-Ranked Preference Optimization (CRPO) to generate and refine audio preference data.

\newpara{Video To Audio Models.}
A key step in automating the Foley process is achieved through Video-To-Audio (V2A) models. Early approaches, such as SpecVQGAN \cite{iashin2021taming}, RegNet \cite{chen2020generating}, and FoleyGAN \cite{xu2024video}, used adversarial training and GAN-based architectures to generate high-quality audio. Diff-Foley \cite{luo2023diff}, a diffusion-based model, introduced CAVP contrastive learning to improve temporal and semantic alignment. Alternatively, Frieren \cite{wang2025frieren}, based on rectified flow, enables efficient audio generation in fewer steps. V-AURA \cite{viertola2024temporally} adopts an autoregressive approach, leveraging the AVCLIP \cite{iashin2024synchformer} representation to extract high-frame-rate temporal and semantic features while bypassing spectrogram conversion. 

Beyond models that require training from scratch, V2A-Mapper \cite{wang2024v2a} and Seeing and hearing \cite{xing2024seeing} employ training-free optimization, utilizing pretrained text-to-audio generators or modality mappers to condition audio generation. While these methods reduce computational cost, they often struggle with fine-grained temporal synchronization, highlighting the ongoing challenge of bridging the gap between video and audio in a seamless and efficient manner.

\newpara{Text and Video To Audio Models.} Text-and-Video-to-Audio (TV2A) models introduce text conditioning to enhance control over synthesized audio. VATT \cite{liu2024tell} leveraged an LLM decoder, functioning as both a video-to-caption model and a video-text-to-audio model. MMAudio \cite{cheng2024taming} and MultiFoley \cite{chen2024video} explicitly trained all three modalities from scratch, achieving state-of-the-art results in signal quality and synchronization. While MMAudio introduced a novel network structure for modality fusion, 
MultiFoley, based on DiT \cite{peebles2023scalable}, leverages multiple conditioning modalities—text, audio, and video—within a single model.
Another approach in TV2A frameworks integrates ControlNet \cite{zhang2023adding} to embed video characteristics into text-to-audio synthesis, as demonstrated by FoleyCrafter \cite{zhang2024foleycrafter} and ReWAS \cite{jeong2024read}. FoleyCrafter extracts frame-based clips as global features in IP-Adapter \cite{ye2023ip} and trains a timestamp detector to identify sound effect occurrences, integrating this information into ControlNet.

ReWAS, a work closely related to ours, divides the training process into two separate stages: a projection network that learns energy features from video using AVCLIP and a ControlNet that utilizes these features to bridge video and audio. Additionally, AudioLDM serves as the foundation for ControlNet, operating on spectrograms and requiring a vocoder to refine the output audio. In the following sections, we demonstrate that our model surpasses ReWAS in semantic and temporal text-based metrics.
 
\section{Conclusion}
\label{sec:conclusion}
In this work, we presented \method, a controllable Automatic Foley Artist designed for the video-and-text-to-audio task. Our model ensures high-quality audio synthesis while maintaining both temporal and semantic alignment with the input video. Guided by text prompts, it allows users to incorporate details beyond what is present in the video or even introduce new creative elements.
This capability enhances flexibility in sound design beyond video-only Foley models. By leveraging the modality adapter, our approach achieves strong performance on a low computational budget. Both objective metrics and human evaluations confirm its effectiveness in generating high-quality, contextually relevant audio.
We believe that further advancements in video feature extraction and T2A model refinement will help address outstanding challenges, such as synthesizing multiple audio sources simultaneously and capturing finer motion details in video.

{
    \small
    \bibliographystyle{ieeenat_fullname}
    \bibliography{references}
}

\appendix
\newpage
\section{Architecture Details}
This section provides additional details about parts of the architecture not covered in the main paper.

\paragraph{AVCLIP preprocessing}

To align the resulting representation with the noised latent space $z_t$, we pre-processed the output of the AVCLIP encoder using two sequential transformation blocks. Each block consists of a fully connected (fc) layer, a ReLU activation, layer normalization, and dropout with $d = 0.1$. In the first block, the fully connected layer maintains the feature dimension at 768, while in the second block, it expands the feature dimension from 768 to 1024.

\paragraph{CLIP preprocessing}
Given a video $V$ of 10 seconds at 25 FPS, we uniformly sampled 5 frames per second and computed their CLIP representations, resulting in feature vectors of shape $(1, 768)$ for each frame. We then applied linear interpolation, producing a signal of shape $(200, 768)$. To ensure alignment with the latent space, we symmetrically padded the signal on both sides so that the 10-second duration corresponds to 216 samples. Next, we processed the signal through a preprocessing block consisting of a fully connected (FC) layer, an ReLU activation, layer normalization, and dropout with $d = 0.1$. In this block, the FC layer preserves the feature dimension of the CLIP encoder output at $768$. We then summed the result with the output of the first AVCLIP processing block and passed it through an additional preprocessing block, where the FC layer expands the feature dimension from $768$ to $1024$, ensuring that the final representation is aligned with the dimensions of $z_t$.

\section{Training Details}
\begin{table}[h]
\centering
\resizebox{\linewidth}{!}{ 
\begin{tabular}{lcccc}
\toprule
\textbf{Model} & \textbf{Steps/Epochs} & \textbf{Hardware} & \textbf{Batch Size} & \textbf{GPU Hours} \\
\midrule
CAFA & 81k steps & A100 (40GB) & 16 & $\sim$48 \\
CAFA-base & 48k steps & A100 (40GB) & 16 & $\sim$24 \\
MMAudio & 300k steps & H100 & - & 304 \\
FoleyCrafter & 164+80 epochs & - & 128 & - \\
Frieren & 2.4M steps & 2 $\times$ RTX 4090 & - & - \\
VATT & ``3 days" of training & A100 (80GB) & - & $\sim$72 \\
MultiFoley & 650k steps & - & 128 & - \\
\bottomrule
\end{tabular}}
\caption{Comparison of training costs across different models. CAFA employs a two-stage training approach (VGGSound followed by VisualSound fine-tuning), while CAFA-base uses single-stage training. FoleyCrafter uses separate semantic (164 epochs) and temporal (80 epochs) training stages. Frieren employs a three-stage approach, and MultiFoley uses a two-stage training method.\label{tab:training_cost}}
\end{table}

\section{User Study Form}
\begin{figure}
    \centering
    \includegraphics[width=1\linewidth]{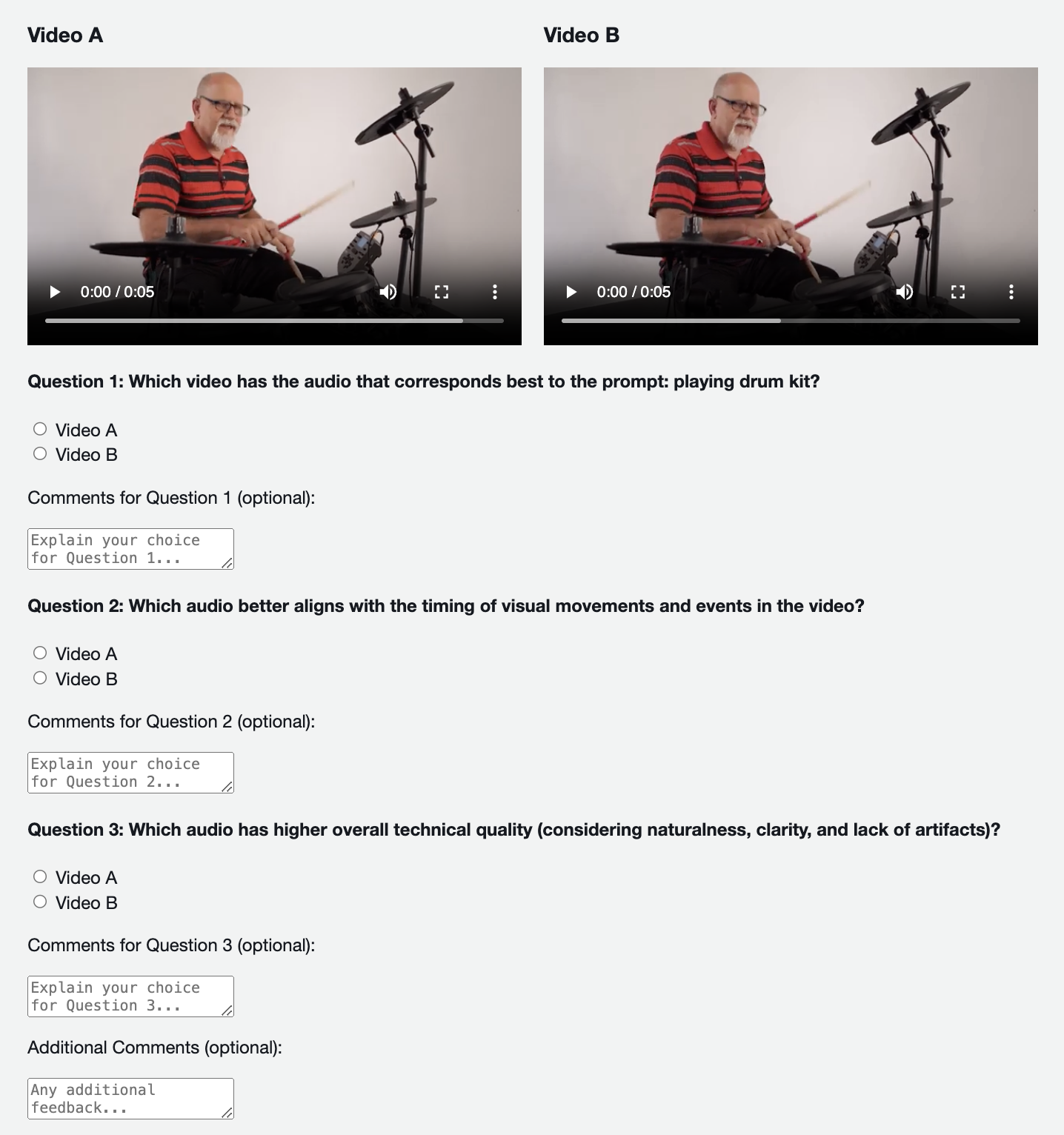}
    \caption{User study form\label{fig:user study}}    
\end{figure}

Figure~\ref{fig:user study} shows the user study interface where participants compared audio outputs from different models, evaluating quality, temporal alignment with video, and adherence to textual prompts.

\section{Additional Figures}
\begin{figure}[h]
    \centering
    \includegraphics[width=0.78\linewidth]{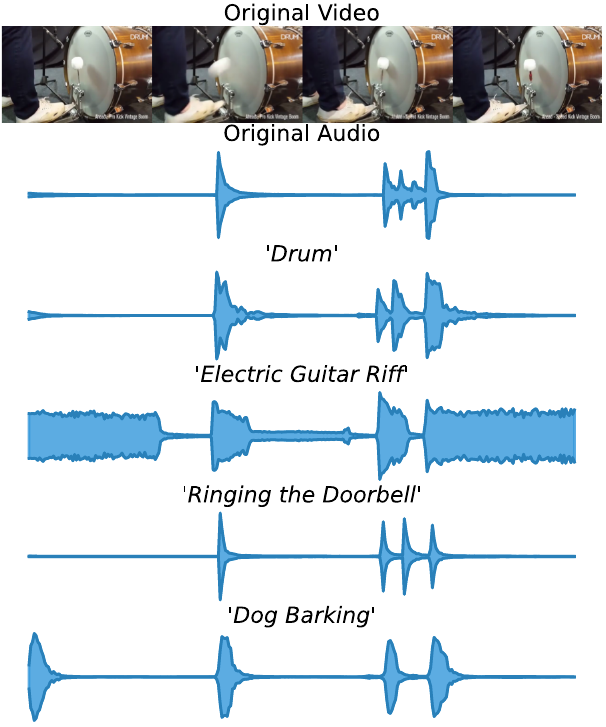}
    \caption{\textbf{CAFA creative control.} We demonstrate our method's ability to generate diverse, high-quality Foley sounds for videos through text prompts, ensuring temporal synchronization between audio and visual elements.}
    \label{fig:drums}
\end{figure}

\end{document}